%% file: main.tex
\documentclass[conference]{IEEEtran}
\IEEEoverridecommandlockouts
\usepackage{cite}
\usepackage{amsmath,amssymb,amsfonts}
\usepackage{algorithmic}
\usepackage{graphicx}
\usepackage{textcomp}
\usepackage{xcolor}
\def\BibTeX{{\rm B\kern-.05em{\sc i\kern-.025em b}\kern-.08em
    T\kern-.1667em\lower.7ex\hbox{E}\kern-.125emX}}
\usepackage[ruled, linesnumbered]{algorithm2e}
\usepackage{algorithm2e,setspace}
\usepackage{booktabs}
\usepackage{multirow}
\newcommand{\ra}[1]{\renewcommand{\arraystretch}{#1}}

\usepackage{array}
\newcolumntype{L}[1]{>{\raggedright\let\newline\\\arraybackslash\hspace{0pt}}m{#1}}
\newcolumntype{C}[1]{>{\centering\let\newline\\\arraybackslash\hspace{0pt}}m{#1}}
\newcolumntype{R}[1]{>{\raggedleft\let\newline\\\arraybackslash\hspace{0pt}}m{#1}}

\makeatletter
 \let\old@ps@headings\ps@headings
 \let\old@ps@IEEEtitlepagestyle\ps@IEEEtitlepagestyle
 \def\confheader#1{%
 \def\ps@IEEEtitlepagestyle{%
 \old@ps@IEEEtitlepagestyle%
 \def\@oddhead{\strut\hfill#1\hfill\strut}%
 \def\@evenhead{\strut\hfill#1\hfill\strut}%
 }%
 \ps@headings%
 }
\makeatother

\confheader{%
\parbox{18cm}{\centering Accepted copy for Publication at the 23rd IEEE International Conference on Industrial Technology (ICIT), 2022 \\ 
Final published version available at: https://doi.org/10.1109/ICIT48603.2022.10002825}}

\begin{document}

\title{Trust your BMS: Designing a Lightweight Authentication Architecture for Industrial Networks}

\author{
\IEEEauthorblockN{Fikret Basic, Christian Steger, Christian Seifert}
\IEEEauthorblockA{\textit{Institute of Technical Informatics} \\
\textit{Graz University of Technology}\\
Graz, Austria \\
\{basic, steger, christian.seifert\}@tugraz.at}
\and
\IEEEauthorblockN{Robert Kofler}
\IEEEauthorblockA{\textit{R\&D Battery Management Systems} \\
\textit{NXP Semiconductors Austria GmbH Co \& KG}\\
Gratkorn, Austria \\
robert.kofler@nxp.com}
}

\maketitle

\begin{abstract}
\input{abstract}

\end{abstract}


\section{Introduction}
\input{introduction}

\section{Background and Related Work}
\input{related_work}

\begin{figure*}[h]
  \centering
  \includegraphics[width=\linewidth]{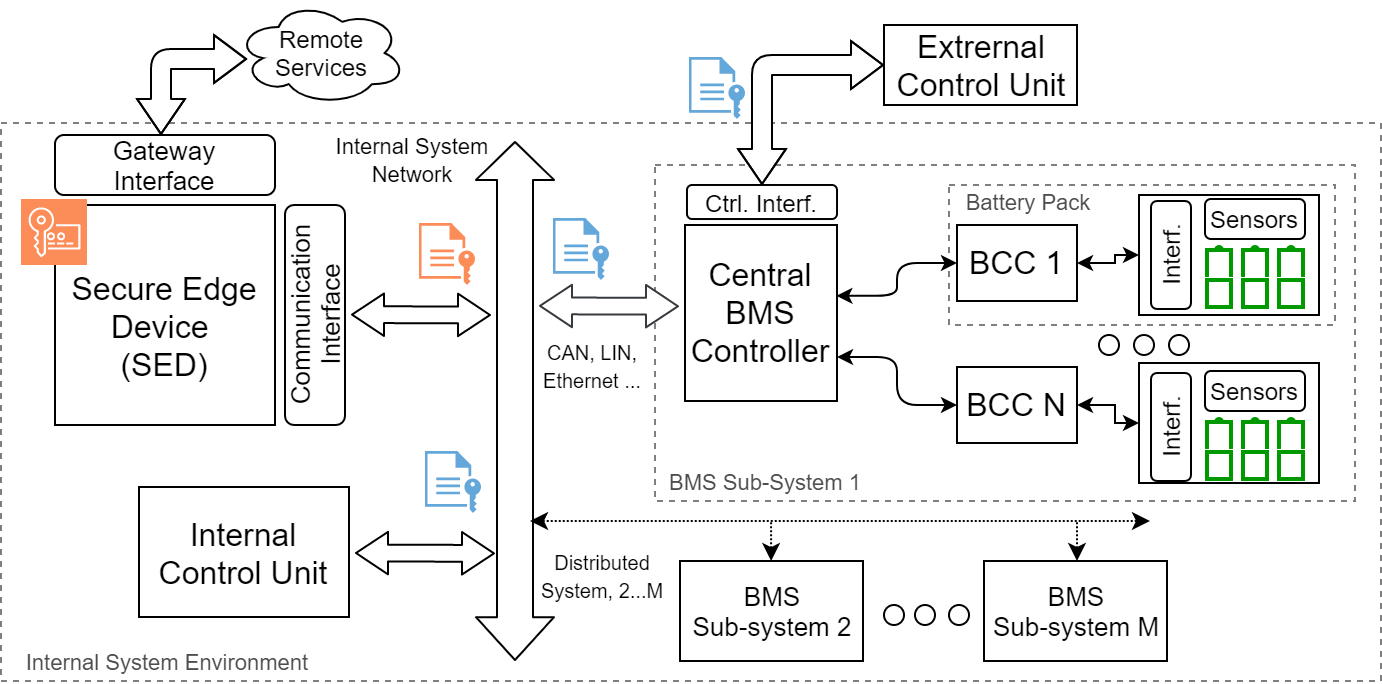}
  \caption{Block demonstration of the proposed security architecture, suggested modules and connections points for the industry systems that contain one or several BMS sub-systems and control devices that interact with them. It showcases potential points of placement regarding SED, BMS and the Control units.}
  \label{fig:bms_sec_arch}
\end{figure*}

\section{Design of a Novel BMS Security Architecture}
\input{design}

\section{Evaluation}
\input{evaluation}

\section{Conclusion and Future Work}
\input{conclusion}

\section*{Acknowledgment}
This project has received funding from the ``EFREtop: Securely Applied Machine Learning - Battery Management Systems'' (Acronym ``SEAMAL BMS'', FFG Nr. 880564).

\bibliographystyle{ieeetr}
\bibliography{references}

\end{document}

%% file: abstract.tex
With the advent of clean energy awareness and systems that rely on extensive battery usage, the community has seen an increased interest in the development of more complex and secure Battery Management Systems (BMS). In particular, the inclusion of BMS in modern complex systems like electric vehicles and power grids has presented a new set of security-related challenges. A concern is shown when BMS are intended to extend their communication with external system networks, as their interaction can leave many backdoors open that potential attackers could exploit. Hence, it is highly desirable to find a general design that can be used for BMS and its system inclusion. In this work, a security architecture solution is proposed intended for the communication between BMS and other system devices. The aim of the proposed architecture is to be easily applicable in different industrial settings and systems, while at the same time keeping the design lightweight in nature.

%% file: introduction.tex
Many systems today rely on large sets of battery cells as power sources. These battery cells are usually packed together in serial or parallel connections.
As the number of these battery cells increases, so does the need for systems that are able to control and automatically respond to different conditions and situations~\cite{Hu2019}. This control is handled through Battery Management Systems (BMS). Today, their usage is rapidly expanding as they are found as part of many different smaller and larger systems. With the increase in the importance of clean energy, BMS are slowly becoming a topic in a broad variety of fields.
Prominent use cases include hybrid and electric vehicles, and smart power grids, where BMS integration is of critical importance for safe and efficient energy control \cite{6532486, 8168251, 7095583}. BMS helps in preventing incidents like the thermal runaway that occurs during the expeditious increase of the battery cell temperature, which would otherwise be difficult to detect~\cite{fireBattery}. 

Each BMS usually consists of a main BMS controller, individual Battery Cell Controllers (BCC), and a battery module that contains battery cells, corresponding sensors, and interfaces.
Traditionally, BMS were deployed as relatively simple sub-systems with limited interaction with the outside components and services. However, when transitioning to larger networks and systems, special attention needs to also be given in the form of protection against malicious attacks~\cite{8813669}. If a device is compromised that is either part of the BMS or the general network, it would give the possibility for a malicious user to mount different attacks. Specifically, an attacker might try to gain direct access to the system, manipulate system data, or even compromise the privacy of a user profile~\cite{Sripad2017, Cheah2019, Kumbhar2018}.

BMS in industrial environments need to be carefully administrated and often require configuration and status updates. These are often done today through external services, such as cloud \cite{book_industrial_cloud} or remote configuration approaches~\cite{8390811}, and a gateway device. However, in internal networks that connect the BMS to the gateway and other components, security is often neglected due to its complexity and design demands. A similar concern has also been addressed in larger smart power grid systems~\cite{6016202, 6032699}. Based on our analysis, we see the following security matters that need to be addressed: (i) configuration data manipulation via exposed interfaces, (ii) industry espionage through Man-in-the-Middle (MitM) attacks, (iii) physical compromise through unauthorized access with a counterfeited or malicious devices.

To address the previously mentioned challenges and security issues, we
consider a design that takes into account the following conditions:
consider the following requirements:
\begin{itemize}
    \item \textit{Portability}: the design needs to allow the exchange and validation of modules between different systems.
    \item \textit{Small footprint}: the implemented security blocks need to be lightweight and not interfere with other operations. 
    \item \textit{Accessibility}: usable between different vendors. 
    \item \textit{Security}: secure under the given operational conditions.
\end{itemize}

We consider the use of the implicit certificates, specifically the Elliptic Curve Qu-Vanstone (ECQV) schema, for establishing fast and efficient network authentication. The use of implicit certificates for in-vehicle authentication has already been previously investigated~\cite{Pullen2019}. However, no specific analysis has yet been conducted related to the use of BMS and its connected services. In this work, we propose an efficient and lightweight design approach for establishing authentication and secure channel communication for BMS and related communication devices. To the best of our knowledge, no other work that investigates this security architectural approach in BMS has been previously proposed.\\
\textbf{Contributions.} Summarized, our main contributions contained in this paper are the following: (i) proposing a BMS secure design architecture for communication with external devices in closed networks, (ii) presenting an authentication protocol based on the implicit certificates, and session key derivation, (iii) using a BMS test device and controllers, we implement the proposed solution and evaluate the process. 

%% file: related_work.tex
\subsection{BMS Security Concepts}
A BMS usually consists of several distinct units. A main BMS controller can communicate with one or many BCCs which in turn can also be connected to one or many battery cell packs. This results in two main security environments that need to be addressed: internal component security, and external service communication. As a relatively new topic that slowly gains interest, research has been mainly focused on the theoretical BMS security models based on the general threat analysis methods \cite{8813669, Kumbhar2018}. While researchers primarily concentrate on the general BMS security models,
Fuchs et al.~\cite{10.1007/978-3-030-54549-9_26} shows a design that uses a Trusted Platform Module (TPM) for establishing a secure communication between BMS and Electric Vehicle Charging Controllers (EVCC).
On the other hand, researchers have also been interested in the BMS cloud environment, proposing design solutions with limited security design considerations \cite{LI2020101557, en11010125}. In this work, we try to bridge the gap between the end point of the BMS controller and direct communication devices to present a design that can be applied for general BMS authentication questions.

\subsection{Authentication Approaches in the Automotive Industry}
Since BMS today play a vital role in the vehicles domain, we have also investigated the State-of-the-Art (SOTA) security architectures inside the vehicle communication environment. Hazem et al.~\cite{Hazem2012LCAPALC} present a protocol for incorporating authentication with the traditional CAN communication protocol. Research conducted by Mundhenk et al.~\cite{7092398} showed an earlier design proposal that includes both the device authentication and secure session establishment between Electronic Control Units (ECUs) in a vehicle. Device authentication is based on combining both asymmetrical and symmetrical crypto approaches and relies on a central security module for control. Similarly, work described in \cite{Pullen2019} extends on the lightweight notion and introduces a general design for in-vehicle authentication of ECUs utilizing Physical Unclonable Functions (PUFs) for the initial device authentication and furthermore implicit certificates for subsequent authentication and key derivation. We do not consider using PUFs for several reason. Mainly, our target features are portability and ease of use of the already established security architectures found in industrial systems and vehicles, especially those that can be established with the verified manufacturers. Furthermore, the PUFs are still largely experimental and based on the recent studies, current implementations have shown vulnerabilities to various threats including machine learning related attacks
~\cite{10.1007/978-3-030-42068-0_3, Dijk_Ruhrmair_2020, 6581556}.

\subsection{Implicit Certificates}
In most modern architectures and networks, systems rely on the use of the explicit certificates usually coupled together with the TLS/SSL for the purpose of authentication and secure communication. 
Research work by Pullen et al.~\cite{Pullen2019} proposed the use of implicit certificates for establishing entity authentication after the initial device authentication. Several other works have also already been conducted handling the implicit certificate implementation, specifically with IoT-related devices \cite{10.1145/3011077.3011108, lightweight_auth_protocol}. Other work includes research conducted in \cite{9500380}, which focuses on the Certificate Transparency (CT) specially aimed to fit the constrained implicit certificate schematic use-cases.
Implicit certificates allow for a lightweight schema without security compromise.

%% file: design.tex
\subsection{Security Requirements}
In an enclosed local network, authentication is an important step usually carried out before other main operations to verify devices that are interconnected. A BMS might need to communicate with additional devices, often to extend the services offered, such as logging and monitoring purposes~\cite{6532486}. Before this communication can take place, the BMS needs to be certain that the device it speaks to is valid and authenticated. Additionally, even if not directly communicated with, every other device inside the network needs to be already authenticated to prevent any kind of sniffing or MitM attacks that could potentially take place \cite{Kumbhar2018}. A potential attacker might either try to attack a BMS for the purpose of reverse engineering and technology exploitation, or data compromise for ransom, frauds, or simply vandalism. 

\subsection{System Architecture}

Our solution is aimed at the modulated BMS topology that uses a central main controller to handle the control of battery packs through BCCs~\cite{10.1007/978-3-030-52794-5_13}. The proposed architecture can also be used for distributed BMS topologies, as each main BMS controller is seen as a separate unit. Through our proposed design the communication to the outside world from the enclosed BMS is only performed through the main control device. This ensures that the main threats, and with that the protection, would be focused on the connection point that the BMS has with external devices. 

The proposed architecture consists of (Fig.~\ref{fig:bms_sec_arch}):
\begin{enumerate}
    \item [$\blacksquare$] \textit{BMS sub-system}: complete modules that include battery controllers and battery packs.
    \item [$\blacksquare$] \textit{Secure Edge Device} (SED): a device that is used both for device authentication and certificate creation and represents the Central Authority (CA) for the local network in this case. It needs to securely handle credential data and fulfill the Common Criteria (CC) conditions.
    \item [$\blacksquare$] \textit{Control Units}: ad-hoc devices attached to the system network, either internally or externally, that want to authenticate a BMS, and need to be authenticated itself.
\end{enumerate}

We assume that the targeted network is closed, i.e., only the SED has access to the outside services (e.g., cloud, monitoring devices). Additionally, any other external communication access (e.g., diagnostic tools) would also need to be verified first as a trusted source by the SED before establishing a connection with other devices in the network.

\subsection{Security Model}
To establish a secure authentication and communication procedure between the BMS and the corresponding devices, a security model was established consisting out of four consecutive steps: (1) fabrication; (2) device authentication; (3) certificate derivation, (4) session communication. Notations used for figures and algorithms are shown in Table~\ref{table:notation}.

The device authentication is proceeded with the \textbf{fabrication} step during which devices are pre-embedded with the necessary security material. This phase is performed only once during the manufacturing stage.

\begin{table}[!ht]
\ra{1.1}
\caption{Notations abbreviation list}
\begin{center}
\begin{tabular}{@{}ll@{}}
\toprule
    Symbol & Description  \\ \midrule
    $N$ & Field key size \\
    $C$ & Random auth. challenge \\
    $key_{auth}$ & Key used for the device auth. \\
    $key_{enc},\; key_{mac}$ & Auth. encryption \& MAC keys \\
    $N_{SED},\; N_{BMS},\; N_{SUM}$ & Auth. random nonces \\
    $ID_{BMS}$ & BMS unique identif. number \\
    $R$ & Response auth. message \\  \midrule
    $t_{BMS},\; k_{BMS}$ & Random private int. values \\
    $P_{BMS}$ & Cert. req. EC point \\
    $U_{BMS},\;S_{BMS}$ & Keys contribution recon. data \\
    $Cert$ & Encoded device certificate \\ \midrule
    $prk_{i},\; pub_{i}$ & Private \& public key of device `i' \\
    $ID_{Sess}$ & Device unique session ID \\
    $chg_{i},\; resp_{i}$ & Auth. challenge \& response \\
    $k_{s}$ & Symmetric session key \\
    \bottomrule
\end{tabular}
\label{table:notation}
\end{center}
\end{table}

\begin{figure}[h]
  \centering
  \includegraphics[width=0.8\linewidth]{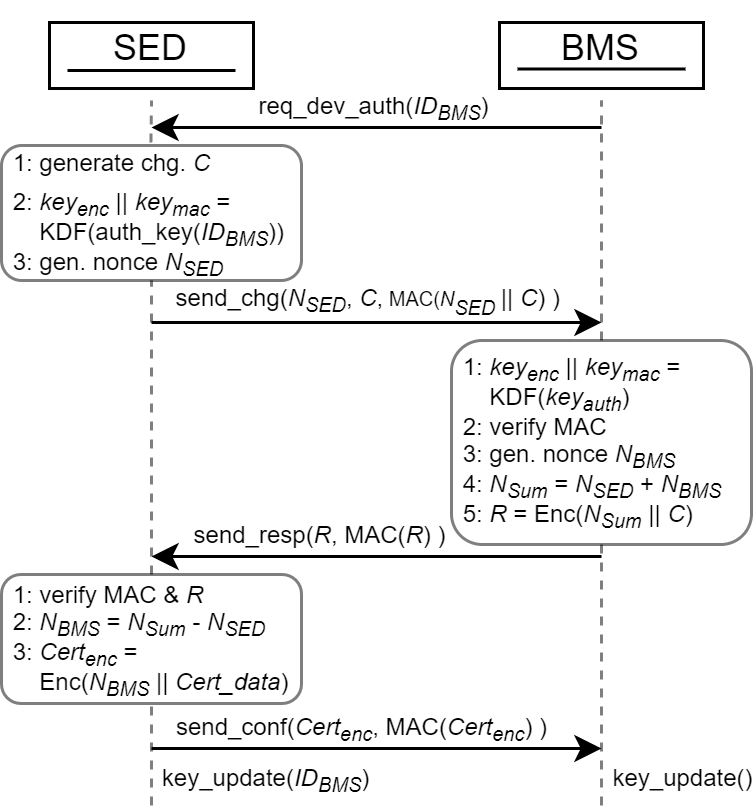}
  \caption{Device authentication process.}
  \label{fig:auth_phase_seq}
\end{figure}

\textbf{Device authentication} step (Fig.~\ref{fig:auth_phase_seq}) uses the Message Authentication Code (MAC) operation for the purpose of handling the authentication procedure. With this, both the BMS and the SED are able to authenticate each other. This process is intended to be run only once when a new device is detected on the network to avoid performance and timing constraints.
The handling is based on the challenge and response mechanism with a \textit{pre-shared key}. Both the SED and the BMS should have a pre-installed secret identifier that can be configured through other secure means \cite{8390811}, with the initial one being established during the fabrication step and used for further key-derivations. Dynamic nonce handling is added for extra protection which includes nonce generation on both entity sides, and the nonce summation and encryption validation~\cite{Pullen2019}. The challenge issued by the SED is concatenated with the random nonces on the BMS side, which is then encrypted and handled with MAC. The extra encryption process helps in preventing potential MitM attacks, particularly replay attacks.

\begin{figure}[!t]
  \centering
  \includegraphics[width=0.8\linewidth]{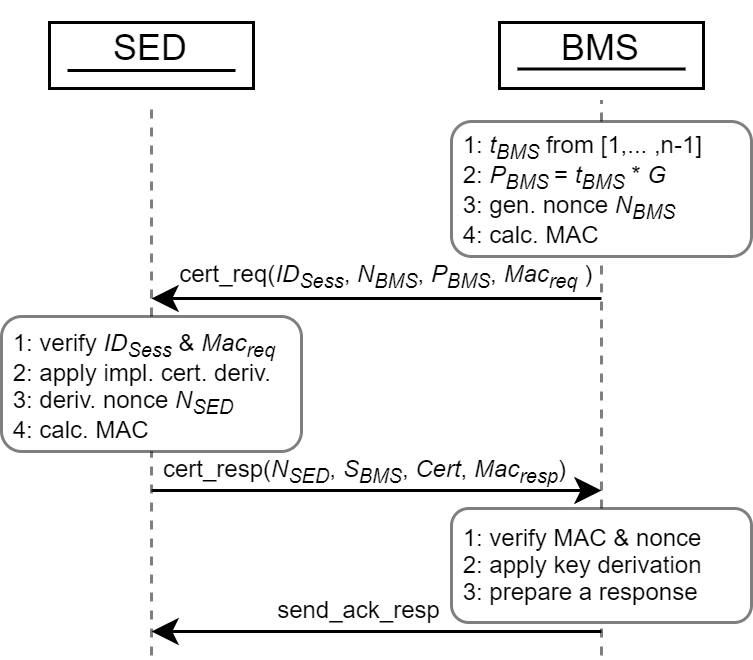}
  \caption{Certificate derivation process.}
  \label{fig:impl_cert_seq}
\end{figure}

\textbf{Certificate derivation} (Fig.~\ref{fig:impl_cert_seq}) follows after the device authentication to complete the configuration process of the newly recognized device. This step is important since the certificates can be afterward used for verification between the BMS and any other device that is part of the network based on the asymmetric cryptography principle. Certificate authentication data is derived and exchanged. To make this possible, during the device authentication, configuration data is sent from the SED to BMS, which contains: a session ID, algorithm identifier (curve, hash), SED's public key, and ID.

The authentication algorithm uses the \textbf{implicit certificates} with the ECQV as the targeted schema for the purpose of deriving and exchanging certificates \cite{Campagna2013}. Based on the proposed ECQV documentation and the ANS.1 format, we decided to use the Minimal Encoding Scheme (MES) without additional extensions for our certificates. The main reason is the smaller certificate sizes, and therefore faster processing than the traditional X.509 format.

The BMS initiates the request for the certificate validation by calculating its necessary construction data, deriving a random nonce, and calculating the MAC value with the previously updated authentication key based on the pre-shared key. Session ID is used to confirm the request. A new session ID is derived on each new device authentication step and is unique for each system device. After verifying the request, the SED will derive the necessary certificate and key construction using Algorithm~\ref{algo:cert_der}. Afterward, a response will be generated and sent back to the BMS where it will first verify the authenticity of the messages based on their MAC and nonce and then proceed with calculating its private and public keys. This key derivation procedure is described by Algorithm~\ref{algo:bms_key_der}.

\begin{algorithm}[!t]
	\caption{SED implicit certificate formulation.}
	\label{algo:cert_der}
	\setstretch{0.9}	
	\footnotesize{
		\SetAlgoLined
		\KwIn{\,\,\,\,$ID_{Sess}$, $P_{BMS}$}
		\KwOut{$S_{BMS}$, $Cert$} 
		Generate $k_{BMS} \in_{R}[1, ..., n-1]$\\
		$U_{BMS} \leftarrow P_{BMS} + k_{BMS}*G$\\
		$Cert \leftarrow Encode(ID_{Sess}, U_{BMS})$\\
		$S_{BMS} \leftarrow (Hash(Cert)*k_{BMS} + prk_{SED}*G)\; mod\; n$\\
		\Return{$S_{BMS},\; Cert$}
	}	
\end{algorithm}

\begin{algorithm}[!t]
	\caption{BMS implicit certificate keys derivation.}
	\label{algo:bms_key_der}
	\setstretch{0.9}	
	\footnotesize{
		\SetAlgoLined
		\KwIn{\,\,\,\,$S_{BMS}$, $Cert$}
		\KwOut{$prk_{BMS}$, $pub_{BMS}$, status} 
		$prk_{BMS} \leftarrow (Hash(Cert)*k_{BMS} + S_{BMS})\; mod\; n$\\
		$pub_{BMS} \leftarrow Hash(Cert)*Decode(Cert) + pub_{SED}$\\
        \eIf { $ pub_{BMS} == prk_{BMS}*G$}{
            \Return{$prk_{BMS},\; pub_{BMS}$}
        }{
            \Return $false$
        }
	}	
\end{algorithm}

\begin{figure}[!ht]
  \centering
  \includegraphics[width=0.8\linewidth]{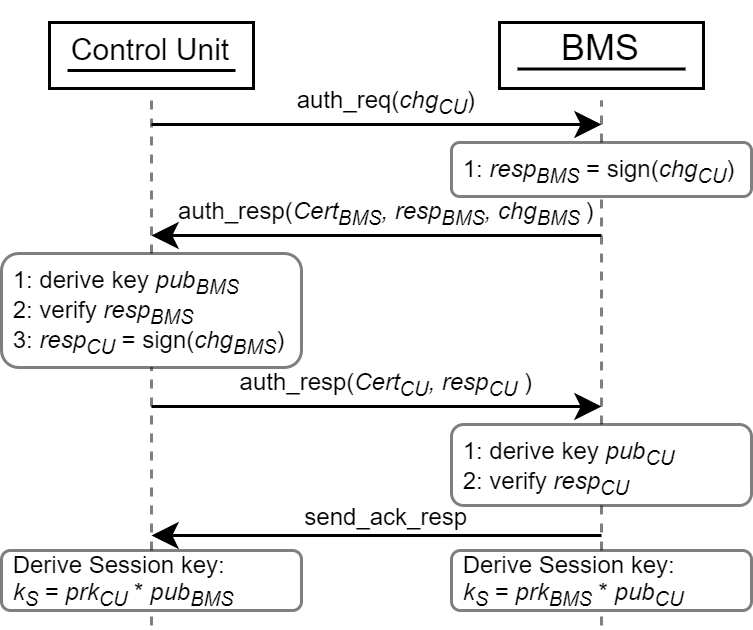}
  \caption{Mutual authentication and session key establishment.}
  \label{fig:session_key_seq}
\end{figure}

\textbf{Session communication} phase (Fig.~\ref{fig:session_key_seq}), is lastly used during a defined session when two devices other than the SED want to mutually authenticate and derive session keys, e.g., the BMS sub-system with a control unit. This phase is coupled together with the certificate derivation for performance reasons since the derived session keys are based on the current public key value and the long-term device private keys~\cite{10.1145/3011077.3011108}. 

\subsection{Discussion on Security Material Updates}
To guarantee a partial \textit{forward secrecy}, i.e., in case older authentication keys are compromised, the keys used in the device authentication phase are updated after each authentication cycle. A Key Derivation Function (KDF) is used to derive new keys based on the previous key and the current request nonce. The initial authentication keys have to be pre-embedded during the fabrication step. With this procedure, even if earlier keys get compromised, the attacker needs to have caught all the previous authentication session interactions and the request nonces to be able to correctly derive the current valid authentication key.

For the certification derivation phase, an open question is made on when should the \textit{re-certification} take place, i.e., when should the new certificates be generated and exchanged. It highly depends on the application's needs, but it is certain to happen at least when certificates expire or during a new system start-up. Otherwise, we propose that the device authentication and re-configuration happen under the following conditions: (i) installation of a new device, (ii) configuration or firmware updates, (iii) changes in the certificate configuration.

%% file: evaluation.tex
\subsection{Prototype Implementation}
\begin{figure}[!t]
  \centering
  \includegraphics[width=0.95\linewidth]{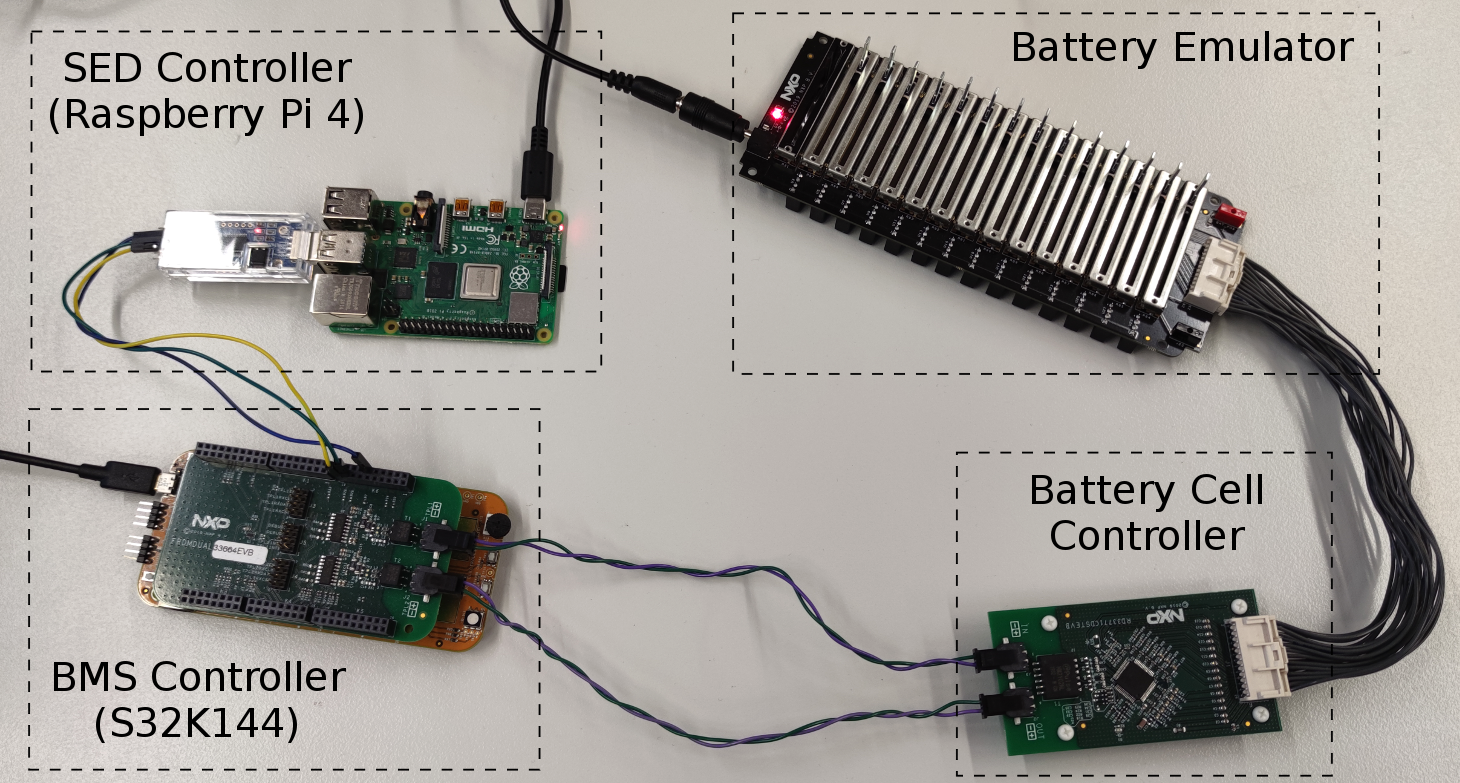}
  \caption{Prototype demonstrator of the proposed security architecture design.}
  \label{fig:bms_proto_setup}
\end{figure}

To evaluate our proposed design approach and analyze its applicability and usability, a prototype test suite was implemented and tested. It was aimed to use higher-grade industry-applicable components with the intention of more closely depicting real-world systems. The test suite consists of full BMS emulation equipment and a Raspberry Pi 4 functioning as a SED. The setup is shown in Fig.~\ref{fig:bms_proto_setup}. 

For the BMS setup, a S32K144 MCU board was used as a central BMS controller. This controller is connected to MC33771C which functions as the BCC. Furthermore, a BATT-14CEMULATOR was used for the emulation of battery cells. The connection between the Raspberry Pi 4 and the BMS controller was established using serial communication with a protocol developed for message handling. SED functionalities have been implemented in Python, with appropriate security handlers using the cryptography library. Encryption is done with the AES-CBC algorithm, where hash (H)MAC is used for the MAC calculations. The lightweight \textit{BearSSL} library was used for the elliptic curve and certificate-related operations. The security software implementation was carefully handled as to still allow the normal flow of the BMS safety control. 

\subsection{Threat Model Analysis}

To test the security feasibility of our design as well as the achieved security level, we have conducted a comprehensive threat model analysis~\cite{Myagmar2005}. The analysis is based on the common attacks indicated by the investigated BMS threat models in \cite{8813669, Sripad2017, Cheah2019, Kumbhar2018}. We assume that the attacker has enough resources and knowledge to launch the potential attacks and that any communication outside of the system is deemed unsafe. We derive the involved Assets~(A), Threats (T), Countermeasures (C), and for threats that are not able to be mitigated, the potential Residual Risks~(R). Afterwards, each threat is classified based on the STRIDE threat categories~\cite{10.5555/579079}, by indicating \underline{S}poofing, \underline{T}ampering, \underline{R}epudiation, \underline{I}nformation Disclosure, \underline{D}enial of Service, and \underline{E}levation of Privilege.

In terms of protection, the following assets need to be secured: \textbf{(A1)} \textit{BMS operational process}: status alerts and adequate safety monitoring, \textbf{(A2)} \textit{Status data}: configuration, raw sensor and derived safety status data, \textbf{(A3)} \textit{Network integrity}: device connectivity, and port access.

The following threats and countermeasures are observed:
\begin{itemize}
    \item \textbf{(T1)}$\langle$S,T,R,I,E$\rangle$ \textit{Malicious update}: attack through configuration data or even code injections. Mitigated by \textbf{(C1)} \textit{Authentication procedure} as proposed in this paper.
    \item \textbf{(T2)}$\langle$I$\rangle$ \textit{Network eavesdrop}: if the attacker gains access to the internal system network. Protected through \textbf{(C1)}, but also \textbf{(C2)} \textit{Encrypted channel}.
    \item \textbf{(T3)}$\langle$T,I$\rangle$ \textit{System data compromise}: affects vulnerable devices that are not properly configured. Either mitigated by \textbf{(C1)} \& \textbf{(C2)}, or not by \textbf{(R1)} \textit{No secure configuration}.
    \item \textbf{(T4)}$\langle$S,T,R,I,D$\rangle$ \textit{Node capturing attacks}: as described in \cite{Porambage2014}. Handled via \textbf{(C3)} \textit{Frequent certificate update control}, and \textbf{(C4)} \textit{Dynamic key updates}.
    \item \textbf{(T5)}$\langle$S,T,R,I,E$\rangle$ \textit{Previous key exposure}: vulnerability depends on the system design and configuration of the updates. Limited protection with \textbf{(C4)} \textit{Forward secrecy}, or, depending on the configuration, \textbf{(R2)} \textit{Updates neglect}.
    \item \textbf{(T6)}$\langle$S,T,R,I,E$\rangle$ \textit{Credentials exposure}: targets either the stored or communicated security material. Mitigated via SED and \textbf{(C5)} \textit{Central access control}.
    \item \textbf{(T7)}$\langle$S,T,R,I$\rangle$ \textit{Counterfeited devices}: fake devices or devices with malicious intent.  Protected with \textbf{(C1)}. 
\end{itemize}

\begin{table}[!ht]
\caption{BMS Time Measurements of Individual Processes}
\begin{center}
\begin{tabular}{@{}clc@{}}
    \toprule
    \multicolumn{2}{c}{BMS (S32K144) Process} & Time (ms)\\
    \midrule
    \multicolumn{1}{c}{\multirow{3}{1.75cm}{Device Authen.}} & 1.1 Prepare req. to SED & $12.6\,\pm\,0.1$ \\
    & 1.3 Handle chg. \& reply & $32.6\,\pm\,0.12$ \\
    & 1.5 Config. \& key update & $5.1\,\pm\,0$ \\
    \midrule
    \multicolumn{1}{c}{\multirow{2}{1.75cm}{Certificate Derivation}} & 2.1 Prepare cert. req. & $651.3\,\pm\,1.3$ \\
    & 2.3 Pub. key calculation & $936.4\,\pm\,5.4$ \\
    \bottomrule
\end{tabular}
\end{center}
\label{table:bms_measurements}
\end{table}

\begin{table}[!ht]
\caption{SED Time Measurements of Individual Processes}
\begin{center}
\begin{tabular}{@{}clc@{}}
    \toprule
    \multicolumn{2}{c}{SED (Rasp. Pi 4) Process} & Time (ms)\\
    \midrule
    \multicolumn{1}{c}{\multirow{2}{1.75cm}{Device Authen.}} & 1.2 Handle req. from BMS & $119.6\,\pm\,3.3$ \\
    & 1.4 Verify resp. from BMS & $7.2\,\pm\,0.2$ \\
    \midrule
    \multicolumn{1}{c}{\multirow{2}{1.75cm}{Certificate Derivation}} & 2.2 Handle req. \& cert. & $238.4\,\pm\,6.4$ \\
    & 2.4 Receive config. Ack & $3.0\,\pm\,0.13$ \\
    \bottomrule
\end{tabular}
\end{center}
\label{table:sed_measurements}
\end{table}

\subsection{Performance Analysis}
To evaluate the application of the design under operational conditions, an execution time analysis has been conducted for critical tasks and steps. Measurements have been run through multiple iterations on both the BMS controller (Table~\ref{table:bms_measurements}) and the SED (Table~\ref{table:sed_measurements}) noting an average value for each vital operation; each noted time includes reading the request, operation handling, and preparing and sending the response.

%% file: conclusion.tex
In this paper, we have presented a novel security architecture solution for BMS in interconnected systems. The design is based on a lightweight solution utilizing efficient symmetric authentication for the initial device verification, and ECQV implicit certificates schema for BMS authentication with internal and external devices and services. The utility of the proposed design was demonstrated through a prototype implementation. To showcase its feasibility, a security evaluation was conducted against common BMS threats, with an additional performance analysis done to investigate the applicability of the design under constrained circumstances.
For future work, we plan to analyse individual authentication mechanisms of distributed battery controllers in enclosed battery packs, and with that to also extend the security handling from the main BMS controller to the other inner modules. Additionally, we would like to exchange our static session key derivation phase with an optimal dynamic key extraction protocol and test its usability.